\documentclass{aa}
\usepackage{graphicx}
\usepackage{txfonts}
\newcommand{\nel}{n_\mathrm{e}}
\newcommand{\Nec}{N_\mathrm{c}}
\newcommand{\nec}{n_\mathrm{c}}
\newcommand{\Filfac}{F_\mathrm{v}}

\newcommand{\avnel}{\langle\nel\rangle}
\newcommand{\ovfilfac}{\overline{f}_\mathrm{v}}
\newcommand{\ovnec}{\overline{n}_\mathrm{c}}

\newcommand{\cmcube}{\,{\rm cm^{-3}}}
\newcommand{\cmsix}{\,{\rm cm^{-6}}}

\newcommand{\pc}{\,{\rm pc}}
\newcommand{\kpc}{\,{\rm kpc}}
\newcommand{\EM}{{\rm EM}}
\newcommand{\RM}{{\rm RM}}
\newcommand{\DM}{{\rm DM}}
\newcommand{\pheins}{\phantom{1}}
\newcommand{\phdrei}{\phantom{111}}
\newcommand{\phelf}{\phantom{11}}
\newcommand{\phc}{\phantom{c}}
\begin{document}
   \title{Densities and filling factors of the DIG in the Solar
          neighbourhood}

%  \subtitle{}

   \author{Elly M. Berkhuijsen and Peter M\"uller}

   \offprints{E.M. Berkhuijsen}

   \institute{Max-Planck-Institut f\"ur Radioastronomie, Auf dem
              H\"ugel 69, 53121 Bonn, Germany\\
              \email{eberkhuijsen@mpifr-bonn.mpg.de};
             \email{peter@mpifr-bonn.mpg.de}
             }

   \date{Received 28 February 2008; accepted 13 July 2008}

% \abstract{}{}{}{}{}
% 5 {} token are mandatory

  \abstract
  {}
{Determination and analysis of electron densities and filling
factors of the diffuse ionized gas (DIG) in the solar neighbourhood.}
{We have combined dispersion measures and emission measures towards 38
pulsars at distances known to better than 50\%, from which we derived
the mean density in clouds, $\Nec$, and their volume filling factor,
$\Filfac$, averaged along the line of sight. The emission measures were
corrected for absorption by dust and contributions from beyond the pulsar
distance. }
{The scale height of the electron layer for our sample is $0.93\pm
0.13\kpc$ and the midplane electron density is $0.023\pm 0.004\cmcube$,
in agreement with earlier results.
The average density along the line of sight is $\avnel = 0.018\pm
0.002\cmcube$ and nearly constant. Since $\avnel = \Filfac \Nec$, an
inverse relationship between $\Filfac$ and $\Nec$ is expected. We find
$\Filfac (\Nec) = (0.011\pm 0.003)\ \Nec^{-1.20\pm 0.13}$, which holds
for the ranges $\Nec = 0.05-1\cmcube$ and $\Filfac = 0.4-0.01$. Near
the Galactic plane the dependence of $\Filfac$ on $\Nec$ is
significantly stronger than away from the plane. $\Filfac$ does not
systematically change along or perpendicular to the Galactic plane, but
the spread about the mean value of $0.08\pm 0.02$ is considerable. The
total pathlength through the ionized regions increases linearly to about
$80\pc$ towards $|z| = 1\kpc$.
} %end results
{Our study of $\Filfac$ and $\Nec$ of the DIG is the first one based on
a sample of pulsars with known distances. We confirm the existence of a
tight, nearly inverse correlation between $\Filfac$ and $\Nec$ in the
DIG. The exact form of this relation depends  on the regions in the
Galaxy probed by the pulsar sample. The inverse $\Filfac$--$\Nec$
relation is consistent with a hierarchical, fractal density
distribution in the DIG caused by turbulence. The observed near
constancy of $\avnel$ then is a signature of fractal structure in the
ionized medium, which is most pronounced outside the thin disk.
} %end conclusions


   \maketitle
%
%________________________________________________________________

\section{Introduction}

The full-sky H$\alpha$ maps of the Milky Way that have recently
become available (Dickinson et al.\ \cite{dickinson+03}; Finkbeiner\
\cite{finkbeiner03}; Bennett et al.\ \cite{bennett+03}; Hinshaw et al.\
\cite{hinshaw+07}) show a spectacular variation in intensity and
structure. The classical \ion{H}{II} regions in the disc are
surrounded by extended, diffuse ionized gas (DIG) visible up to high
latitudes, and supernova explosions, shocks and turbulence in the
interstellar medium (ISM) have produced a wealth of cloud shapes,
filaments, shells and voids.

Madsen et al. (\cite{madsen+06}) found the variations in structure to
be accompanied by variations in electron temperature. They derived a
mean temprature of the DIG of about $8000\,$K with strong differences
between lines of sight.

A number of authors derived a scale height of the DIG in the solar
neighbourhood of about $1\kpc$ from the increase of the dispersion
measures of pulsars with increasing distance to the Galactic plane
(Reynolds\ \cite{reynolds91b}; Bhattacharia \& Verbunt\
\cite{bhatta+91}; Nordgren et al.\ \cite{nordgren+92}; Gomez et
al.\ \cite{gomez+01}; Cordes \& Lazio\ \cite{cordes+lazio03}). Using
dispersion measures towards 4 pulsars in globular clusters, Reynolds
(\cite{reynolds91a}) obtained a filling fraction of the DIG of $\la 0.2$
through the full layer and a mean density in clouds of about
$0.08\cmcube$. This was recently confirmed by Berkhuijsen et al.
(\cite{elly+06}) from a much larger pulsar sample. These authors also
found an increase of the filling factor with distance from the plane.
Already the scanty data that were available to Kulkarni \& Heiles
(\cite{kulkarni+88}) had indicated such an increase.

Turbulence causing hierarchical, fractal structure leads to an inverse
relationship between the mean filling factor along the line of sight,
$\Filfac$, and the mean density in ionized regions, $\Nec$
(Fleck\ \cite{fleck96}; Elmegreen\ \cite{elmegreen98},
\cite{elmegreen99}). Pynzar (\cite{pynzar93}), who
was the first to investigate this relationship, derived $\Filfac
\propto \Nec^{-0.7}$ for a combination of DIG and \ion{H}{II} regions,
which was confirmed by Berkhuijsen (\cite{elly98}). In an extensive
study, Berkhuijsen et al. (\cite{elly+06}, hereafter called BMM)
obtained $\Filfac \propto \Nec^{-1.07\pm 0.03}$ for the DIG from a
sample of 157 pulsars at latitudes $|b|> 5\degr$. They used dispersion
measures from the catalogue of Hobbs \& Manchester
(\cite{hobbs+03}; see also Manchester et al.\
\cite{manchester+05}), emission measures from the WHAM survey (Haffner
et al.\ \cite{haffner+03}), corrected for absorption (Diplas \& Savage\
\cite{diplas+savage94}) as well as for contributions from beyond the
pulsars, and pulsar distances from the model of the electron
distribution in the Galaxy of Cordes \& Lazio (\cite{cordes+lazio02}).
The statistical error in pulsar distances $<3\kpc$, where 75\% of the
pulsars in the sample are located, is about 25\% which is much smaller
than the intrinsic spread in the dispersion measures and emission
measures. Therefore, random errors in the model distances will not
have influenced the statistical results. However, the distances from
the model may also contain a systematic error. Based on these
distances, the radial distribution of pulsars has a maximum at
$R=3.5\kpc$ from the Galactic centre (Lorimer\ \cite{lorimer04};
Yusifov \& K\"u\c{c}\"uk\ \cite{yusifov+04}), whereas other population~I
objects (H2, \ion{H}{I}, \ion{H}{II} regions) peak near $R=5\kpc$. The
smoothness of the electron density model may lead to pulsar distances
that are too large.

So it is important to check the results of BMM for pulsars with distances
derived from observations. Since BMM did their work, the sample
of pulsars with measured distances has increased considerably. Our new
analysis is based on 38 pulsars, the distances of which are known to
better than 50\%. A further decrease of the constraint on the distance
errror would make the sample too small for a statistical analysis.
We also used an improved absorption correction to the emission measure
of each pulsar instead of the statistical correction applied by BMM.

\section{Basics and data}

\subsection{Basic relations}

The expressions for dispersion measure, $\DM$, and emission measure,
$\EM$, towards a pulsar at distance $D$ (in pc) can be written in
various ways:

\begin{equation}    \label{eq-dm}
\DM = \int^\mathrm{D}_0 \nel(l)dl = \avnel D = \Nec \Filfac D =
 \Nec L_\mathrm{e}\ ,
\end{equation}
\begin{equation}   \label{eq-em}
\EM  = \int^\mathrm{D}_0  \nel^2(l)dl = \langle\nel^2\rangle D =
  \Nec^2 \Filfac D = \Nec^2 L_\mathrm{e}\ ,
\end{equation}
where $\nel(l)$ (in $\mbox{cm}^{-3}$) is the local electron density
at point $l$ along the line of sight, $L_\mathrm{e}$ (in pc) the
total pathlength through the regions containing free electrons
(clouds, clumps) and $\Nec$ (in $\mbox{cm}^{-3}$) the average
density in these regions; between clouds the electron density is
assumed to be negligible (see Fig.~1 in BMM). Furthermore, $\avnel$ and
$\langle\nel^2\rangle$ are averages along $D$, and $\Filfac =
L_\mathrm{e}/D$ is the fraction of the line of sight in clouds, which
approximates the volume filling factor $\Filfac$ if there are several
clouds along the line of sight (BMM). Note that the third equality in
eq.~(\ref{eq-em}) is only valid when the average density of every cloud
$\nec$ along the line of sight is the same: then $\langle\nel^2\rangle
= \langle\nec^2\rangle \Filfac = \Nec^2 \Filfac$. Thus $\Nec$ and
$\Filfac$ are approximations of the true average density in clouds and
their filling factor.

Combining eqs.~(\ref{eq-dm})\footnote{Note that we write $\Filfac$ and
$\Nec$ where BMM used $\ovfilfac$ and $\ovnec$} and (\ref{eq-em}) we
have

\begin{equation}   \label{eq-Nc}
\Nec = \EM/\DM\ ,
\end{equation}
\begin{equation}   \label{eq-Fd}
\Filfac = \DM^2 / \EM\, D\ .
\end{equation}
The derived quantities $\Nec$ and $\Filfac$ are connected by the simple
relation $\avnel = \Filfac \Nec$ following from the third equality in
eq.~(\ref{eq-dm}). As the mean electron density in clouds will not be
constant, our estimate of $\Nec$ will be too high because of the
$\nel^2$ dependence of $\EM$. Recently, Hill et al. (\cite{hill+08})
found that the ratio between the most probable density and $\Nec$ varies
between 0.95 and 0.4 in the mildly supersonic cases of their MHD
simulations (scaled to observations). Thus, our values of $\Nec$ may be
too high by a factor $\la 2$; the corresponding values of $\Filfac$ will
then be too low by the same factor.

\subsection{The data}

We collected a sample of 78 pulsars with distances known to better
than 50\% from the literature. This error is negligible compared to
the large intrinsic spread in $\DM$ (factor of 2) and $\EM$ (factor
of 4). The sample consists of 52 pulsars from the list of Gomez et
al. (\cite{gomez+01}), where we used more recent distances when
available, and 26 pulsars with parallactic distances taken from
Hobbs et al. (\cite{hobbs+05}, and references therein).

The dispersion measures came from the catalogue of Manchester et al.
(\cite{manchester+05}). To find the emission measures in the pulsar
directions we used the full-sky H$\alpha$ map of Finkbeiner
(\cite{finkbeiner03}) corrected for extinction as described by
Dickinson et al. (\cite{dickinson+03}), who argued that only one
third of the dust in the line of sight effectively absorbs H$\alpha$
emission. The extinction-corrected emission measure is

\begin{displaymath}
\EM_\mathrm{c} = \EM \times 0.34\
A\quad  \hbox{with}\  A = \exp \left( 2.4\,
 E(B-V)\right)\ ,
\end{displaymath}
where $\EM$ is the observed emission measure and $A$ the correction
recommended by Finkbeiner (\cite{finkbeiner03}). Corrections of more
than 1 magnitude are uncertain and cannot be used. Therefore we had
to remove 38 pulsars from the sample, most of which are at latitudes
$|b| < 5\degr$. We also removed 2 pulsars with high $\EM$ and $\DM$
(J0613$-$0200 and J1807+0943) indicating \ion{H}{II} regions on
their lines of sight. Thus for the analysis a sample of 38 pulsars
is available, for 23 of which a parallactic distance is known. The
errors in the distances to 28 pulsars are smaller than 20\% ($\Delta
D/D < 0.2)$ and those in the distances to the other 10 pulsars are
less than 50\% ($0.2 < \Delta D/D < 0.5$). As 21 pulsars are located
at D $<2\kpc$, the sample is heavily weighted to the solar
neighbourhood; 13 pulsars are in globular clusters at distances
$>2\kpc$. The main parameters of the 38 pulsars in the sample are
listed in Table~\ref{tab:1}, which also gives the references to the
distances used. The sightlines to these pulsars probe the DIG.

\begin{table*}
\caption{Parameters of the 38 pulsars in the final sample}
\label{tab:1}
\centering
\begin{tabular}{c r r c c c c r r c c}
\hline\hline
  PULSAR      &LONG     &LAT       &$\DM^a$         &\multicolumn{2}{c}{\hspace{0.4cm} $D$~~$\pm$~~$\Delta D^b$}
    &REF   &$\EM_\mathrm{c}$   &$\EM_\mathrm{p}$   &$\Nec$   &$\Filfac$ \\
  Jname       &$[\degr]$ &$[\degr]$ &$[\cmcube\pc]$ &\multicolumn{2}{c}{\hspace{0.4cm} [pc]}
    &      &\multicolumn{2}{c}{$[\cmsix\pc]$}       &$[\cmcube]$  \\
\hline
J0437$-$4715\phc  &253.40  &$-$42.00  &\phelf 2.65    &\phelf 159   &\phdrei 5    &\pheins 1  & 1.8   &0.4   &0.137   &0.122 \\
J0814+7429\phc    &140.00  &   31.62  &\phelf 6.12    &\phelf 433   &\phdrei 8    &\pheins 5  & 1.9   &0.7   &0.116   &0.122 \\
J0826+2637\phc    &197.00  &   31.70  &\pheins 19.45  &\phelf 360   &\phelf 80    &\pheins 6  & 4.8   &1.6   &0.081   &0.665 \\
J0953+0755\phc    &228.91  &   43.70  &\phelf 2.96    &\phelf 262   &\phdrei 5    &\pheins 5  & 3.2   &1.0   &0.351   &0.032 \\
J1136+1551\phc    &241.91  &   69.19  &\phelf 4.86    &\phelf 350   &\phelf 20    &\pheins 5  & 1.5   &0.8   &0.155   &0.090 \\
J1239+2453\phc    &252.45  &   86.54  &\phelf 9.24    &\phelf 850   &\phelf 60    &\pheins 5  & 1.1   &0.9   &0.100   &0.108 \\
J1456$-$6843\phc  &313.90  & $-$8.50  &\phelf 8.60    &\phelf 450   &\phelf 60    &\pheins 8  &32.4   &4.3   &0.497   &0.038 \\
J1932+1059\phc    & 47.38  & $-$3.88  &\phelf 3.18    &\phelf 330   &\phelf 10    &\pheins 5  &18.5   &0.9   &0.269   &0.036 \\
J2018+2839\phc    & 68.10  & $-$3.98  &\pheins 14.18  &\phelf 950   &\phelf 90    &\pheins 5  &35.0   &4.6   &0.323   &0.046 \\
J2022+2854\phc    & 68.86  & $-$4.67  &\pheins 24.64  &\pheins 2300 &\pheins 800  &\pheins 5  &38.5   &12.6  &0.513   &0.021 \\
J2022+5154\phc    & 87.86  &    8.38  &\pheins 22.65  &\pheins 1900 &\pheins 250  &\pheins 5  &12.6   &5.6   &0.247   &0.048 \\
J0922+0638\phc    &225.42  &   36.39  &\pheins 27.27  &\pheins 1210 &\phelf 90    &\pheins 7  & 7.3   &5.7   &0.209   &0.108 \\
J1537+1155\phc    & 19.85  &   48.30  &\pheins 11.61  &\pheins 1080 &\pheins 150  &\pheins 1  & 2.7   &2.2   &0.192   &0.056 \\
J1713+0747\phc    & 28.75  &   25.22  &\pheins 15.99  &\phelf 910   &\phelf 80    &\pheins 1  & 4.9   &2.8   &0.172   &0.102 \\
J1744$-$1134\phc  & 14.79  &    9.18  &\phelf 3.14    &\phelf 470   &\phelf 95    &\pheins 1  &21.4   &3.2   &1.005   &0.007 \\
J1909$-$3744\phc  &359.73  &$-$19.60  &\pheins 10.39  &\pheins 1140 &\phelf 50    &\pheins 1  & 8.1   &4.5   &0.431   &0.021 \\
J0659+1414\phc    &201.11  &    8.26  &\pheins 13.98  &\phelf 290   &\phelf 30    &\pheins 4  &14.0   &1.2   &0.085   &0.567 \\
J2145$-$0750\phc  & 47.78  &$-$42.08  &\phelf 9.00    &\phelf 500   &\pheins 200  &\pheins 9  & 1.7   &0.9   &0.098   &0.184 \\
J1022+1001\phc    &231.79  &   51.10  &\pheins 10.25  &\phelf 400   &\pheins 145  &\pheins 1  & 2.7   &1.3   &0.129   &0.198 \\
J1024$-$0719\phc  &251.70  &   40.52  &\phelf 6.48    &\phelf 520   &\pheins 270  &\pheins 1  & 2.3   &1.2   &0.183   &0.068 \\
J2124$-$3358\phc  & 10.93  &$-$45.44  &\phelf 4.60    &\phelf 250   &\pheins 165  &\pheins 1  & 2.8   &0.9   &0.189   &0.098 \\
J0030+0451\phc    &113.14  &$-$57.61  &\phelf 4.33    &\phelf 300   &\phelf 90    &\pheins 2  & 1.0   &0.4   &0.099   &0.146 \\
J1509+5531\phc    & 91.32  &   52.29  &\pheins 19.61  &\pheins 2370 &\pheins 220  &\pheins 3  & 1.0   &1.0   &0.049   &0.169 \\
J0534+2200\phc    &184.56  & $-$5.78  &\pheins 56.79  &\pheins 2000 &\phelf 75    &11         &23.3   &8.1   &0.143   &0.198 \\
J1740$-$5340c     &338.20  &$-$11.90  &\pheins 71.80  &\pheins 2300 &\phelf 20    &11         &37.8   &24.0  &0.335   &0.093 \\
J0141+6009\phc    &129.10  & $-$2.10  &\pheins 34.80  &\pheins 2750 &\pheins 150  &11         &29.0   &5.6   &0.161   &0.079 \\
J1910$-$5959c     &336.50  &$-$25.60  &\pheins 33.68  &\pheins 4000 &\pheins 250  &11         & 3.2   &3.1   &0.092   &0.091 \\
J1824$-$2452c     &  7.80  & $-$5.58  &119.86         &\pheins 5700 &\phelf 750   &11         &76.6   &53.1  &0.443   &0.047 \\
J1701$-$3006c     &353.60  &    7.30  &114.56         &\pheins 6900 &\pheins 900  &11         &59.4   &50.2  &0.438   &0.038 \\
J1911+0101c       & 36.11  & $-$3.92  &202.68         &\pheins 9500 &\pheins 800  &12         &39.2   &29.3  &0.145   &0.147 \\
J1823$-$3021c     &  2.79  & $-$7.91  &\pheins 86.84  &\pheins 8000 &\pheins 800  &11         &26.8   &24.2  &0.279   &0.039 \\
J1804$-$0735c     & 20.79  &    6.77  &186.32         &\pheins 8400 &1900         &11         &35.5   &31.2  &0.168   &0.132 \\
J1748$-$2021c     &  7.73  &    3.80  &220.40         &\pheins 8400 &2100         &11         &88.7   &61.6  &0.279   &0.094 \\
J1721$-$1936c     &  4.86  &    9.74  &\pheins 75.70  &\pheins 8600 &1100         &11         &21.4   &20.4  &0.270   &0.033 \\
J0024$-$7204c     &305.92  &$-$44.89  &\pheins 24.60  &\pheins 4500 &\pheins 250    &11 & 3.6   &...   &...     &...  \\
J1518+0205c       &  3.8   &   46.80  &\pheins 29.47  &\pheins 7500 &\pheins 400    &11 & 2.1   &...   &...     &...   \\
J1641+3627c       & 59.00  &   40.91  &\pheins 30.36  &\pheins 7700 &\pheins 380    &11 & 1.8   &...   &...     &...  \\
J2129+1210c       & 65.01  &$-$27.31  &\pheins 67.31  &10300        &\pheins 650    &11 & 5.0   &...   &...     &...  \\
\hline
\noalign{\smallskip}
\multicolumn{11}{l}{{\it a)} Hobbs \& Manchester (\cite{hobbs+03}):
{\tt www.atnf.csiro.au/research/pulsar/psrcat} }\\
\multicolumn{11}{l}{{\it b)} Mean of absolute values of positive and
negative error.} \\
\multicolumn{11}{l}{{\it c)} In globular cluster. The last four
pulsars are at $|z|> 3\kpc$, and only used in the determination of the } \\
\multicolumn{11}{l}{electron scale height (see Fig.~\ref{fig:1}). } \\
\multicolumn{11}{l}{References. (1) Hotan et al.\ \cite{hotan+06}; (2)
Lommen et al.\ \cite{lommen+06}; (3) Chatterjee et al.\
\cite{chatter+05}; (4) Brisken et al.\ \cite{brisken+03}; } \\
\multicolumn{11}{l}{(5) Brisken et al.\ \cite{brisken+02}; (6) Gwinn et
al.\ \cite{gwinn+86}; (7) Chatterjee et al.\ \cite{chatter+01}; (8) Bailes et al.\
\cite{bailes+90}; } \\
\multicolumn{11}{l}{(9) L\"ohmer et al.\ \cite{loehmer+04}; (10) Stairs
et al.\ \cite{stairs+98}; (11) Gomez et al.\ \cite{gomez+01}; (12)
Heitsch \& Richtler\ \cite{heitsch+99}. }\\
\end{tabular}
\end{table*}
%--------------------------------------------------------------------------------

In the direction of nearby pulsars and of pulsars near the
Galactic plane a significant amount of the emission measure
originates beyond the pulsar (see Fig.~5 in BMM).
Reynolds (\cite{reynolds97}) and Haffner et al. (\cite{haffner+98})
have shown that the vertical electron distribution derived from
H$\alpha$ observations is fairly well described by an
exponential. Then the emission measure up to the pulsar,
$\EM_\mathrm{p}$, is obtained from

\begin{equation}   \label{eq-emp}   %%eq-5
\EM_\mathrm{p} \sin|b|(z) = \EM_\mathrm{c} \sin|b|
\left( 1-\exp \left( -|Z_\mathrm{p}|/h_\mathrm{e}\right) \right) \ ,
\end{equation}
where $Z_\mathrm{p}$ is the distance of the pulsar to the Galactic
plane and $h_\mathrm{e}$ is the scale height of $\nel^2(z)$.
Deviations from a smooth, plane-parallel layer will cause scatter
in $\EM_\mathrm{p}$. The mean of $\EM_\mathrm{c} \sin|b|$ for our
sample is $2.8\pm 0.3\cmsix \pc$, which is the extinction-corrected
emission measure through the full layer. We derive
the scale height $h_\mathrm{e}$ in the next section.

\section{Scale height of electrons}

\begin{figure} %fig 1
\includegraphics[bb = 98 61 549 669,angle=270,width=8.8cm]{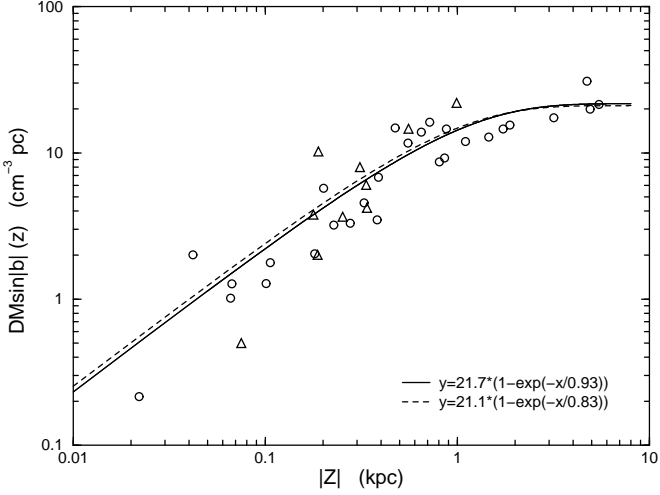}
\caption{Dependence of the dispersion measure perpendicular to the
Galactic plane, $\DM \sin|b|$, on distance to the plane, $|z|$, of
the final sample of 38 pulsars. Circles: pulsars with
$\Delta D/D < 0.2$; triangles: pulsars with $0.2 < \Delta D/D <
0.5$. Full line: the two-parameter fit to the data; dashed line: fit
to the original sample of 78 pulsars, shown for comparison.}
\label{fig:1}
\end{figure}

We first estimated the scale height $h_\mathrm{e}$ of $\nel^2(z)$ using
Eqs.~12 and 13 of BMM, expecting that also for our sample
the average filling factor $\Filfac (z)$ would increase with $|z|$.
This yielded $280\pc < h_\mathrm{e} < 510\pc$ with a best value of
$390\pc$, giving the observed maximum values of $\DM\sin|b| =
22\pm 2\cmcube\pc$ (see Fig.~\ref{fig:1}) and $\EM_\mathrm{p} \sin|b| =
2.8\pm 0.3\cmsix\pc$. However, for this entire range of $h_\mathrm{e}$,
$\Filfac (z)$ appeared to be essentially constant for our sample (see
Fig.~\ref{fig:4}d), thus also the local filling factor $f(z)$ is
$\simeq$ constant. As is easily seen, this indicates that the scale
height of $\nel^2(z)$ is about half that of $\nel (z)$.

Following BMM, we describe the dispersion measure perpendicular
to the Galactic plane as

\begin{eqnarray}   \label{eq-6}
\DM \sin|b|(z) &= & \int_0^{|Z_\mathrm{p}|} \nel(z)dz
                  = \int_0^{|Z_\mathrm{p}|} f(z) \nec(z)dz \nonumber \\
   &= &f_0 n_\mathrm{c0} h \left( 1-\exp\left(
       -|Z_\mathrm{p}|/h\right) \right) \ ,
\end{eqnarray}
where we assume that $f(z)=f_0$ and the local density in clouds
$\nec(z)$ is an exponential with midplane value $n_\mathrm{c0}$ and
scale height $h$; $Z_\mathrm{p}$ is the distance of the pulsar to the
midplane.

Similarly, the emission measure perpendicular to the Galactic
plane is

\begin{eqnarray}    \label{eq-7}
\EM_\mathrm{p} \sin|b|(z) &= & \int_0^{|Z_\mathrm{p}|} \nel^2 (z)dz
           = \int_0^{|Z_\mathrm{p}|} f(z) \nec^2 (z)dz \nonumber \\
   & = &f_0 n_\mathrm{c0}^2 h \left( 1-\exp\left(
-2|Z_\mathrm{p}|/h\right)\right)/2 \ ,
\end{eqnarray}
where the scale height $h_\mathrm{e}$ in eq.~(\ref{eq-emp}) is
equal\footnote{Note that BMM defined $h$ as the scale height of
$\nel^2(z)$, thus their $h$ equals our $h_\mathrm{e} = h/2$.} to
$h/2$.

We can derive the maximum of $\DM \sin|b|$ and the scale height $h$
from the distribution of $\DM \sin|b|(z)$ of our sample of 38 pulsars
shown in Fig.~\ref{fig:1}. A two-parameter fit yielded $f_0
n_\mathrm{c0} h = 21.7\pm 1.5\cmcube\pc$ and $h = 0.93\pm 0.13\kpc$,
giving $\nel(0) = f_0 n_\mathrm{c0} = 0.023\pm 0.004\cmcube$.
As the maximum of $\EM_\mathrm{p} \sin|b|$ equals
$\EM_\mathrm{c} \sin|b|$, we have $\nel^2(0)h/2 = f_0 n_\mathrm{c0}^2
h/2 = 2.8\pm 0.3\cmsix\pc$ (see Sect.~2); hence $\nel^2(0) = 0.0060
\pm 0.0011\cmsix$, $n_\mathrm{c0} = 0.26\pm 0.03\cmcube$ and $f_0 =
0.09\pm 0.02$.

\begin{figure} %fig 2
\includegraphics[bb = 98 61 549 669,angle=270,width=8.8cm]{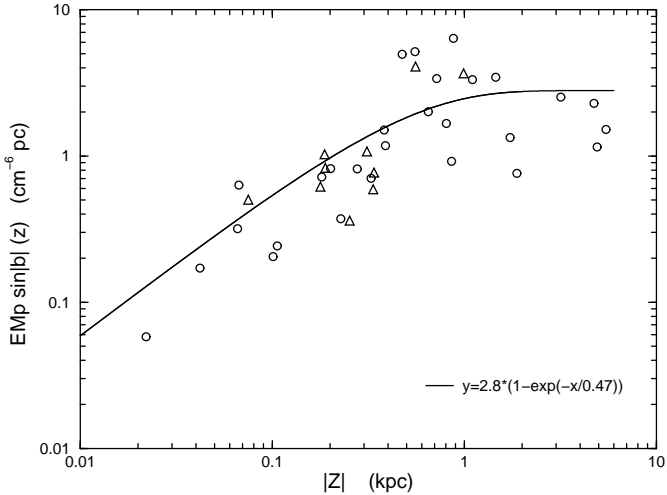}
\caption{Dependence of the corrected emission measure perpendicular to
the Galactic plane, $\EM_\mathrm{p}sin|b|$, on distance to the plane,
$|z|$, of the final sample of 38 pulsars. Circles: pulsars with
$\Delta D/D < 0.2$; triangles: pulsars with $0.2 < \Delta D/D <0.5$.
Full line: dependence expected for the scale height of $0.47\kpc$ used
in eq.~(\ref{eq-emp}).}
\label{fig:2}
\end{figure}

We then analyzed the data in our sample using $h_\mathrm{e} = h/2 =
0.47\kpc$ in eq.~(\ref{eq-emp}), leaving out 4 pulsars in globular
clusters at $|z| > 3\kpc$. As only a small part of their lines of sight
passes through the electron layer, the mean electron densities along
the line of sight are unrealistically low. Figure~\ref{fig:2} shows the
distribution of $EM_\mathrm{p}\sin|b|(z)$ for all pulsars. The
steady increase of $EM_\mathrm{p} \sin|b|$ with $|z|$ agrees well with
the expected variation (full line), but the scatter is larger than in
Fig.~\ref{fig:1}.

\section{Results}

In this section we investigate the dependencies of $\avnel$,
$\langle\nel^2\rangle$, $\Nec$ and $\Filfac$ on $|z|$, and the
relationship between $\Filfac$ and $\Nec$. The statistical treatment is
the same as used by BMM and we refer to that paper for details. The
results are given in Table~\ref{tab:2}.

\begin{figure} %fig 3
\includegraphics[bb = 98 61 549 677,angle=270,width=8.8cm]{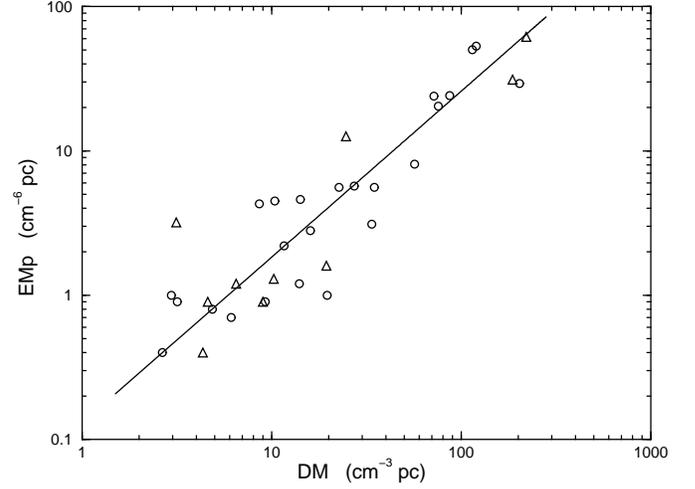}
\caption{Comparison of corrected emission measure, $\EM_\mathrm{p}$,
and observed dispersion measure, $\DM$. Circles: pulsars with
$\Delta D/D < 0.2$; triangles: pulsars with $0.2 < \Delta D/D <
0.5$. Full line: bisector fit given in Table~\ref{tab:2}. }
\label{fig:3}
\end{figure}

We first show in Fig.~\ref{fig:3} that $\EM_\mathrm{p}$ is well
correlated with $\DM$: the bisector fit yields $\EM_\mathrm{p} = (0.13
\pm 0.04) \DM^{1.15\pm 0.07}$ with very high significance. This indicates
that both quantities probe the same ionized regions along the
line of sight, although the beams used for the H$\alpha$ observations
(up to 1\degr ) also sample regions located around the single sightline
to the pulsar. Variations in electron density across the beam will
contribute to the scatter in Fig.~\ref{fig:3}. The components of
$\EM_\mathrm{p}$ and $\DM$ along $|z|$ correlate less well than
those along the Galactic plane (see Table~\ref{tab:2}), because of the
different scale heights of $\nel^2(z)$ and $\nel(z)$.

\begin{table*}
\caption{Statistical relationships derived for the sample of 34 pulsars}
\label{tab:2}
\centering
\begin{tabular}{l l l r@{$\pm$}l r@{$\pm$}l c c}
\hline\hline
$X$  &$Y$  &Fit  &\multicolumn{2}{c}{$a$}  &\multicolumn{2}{c}{$b$}  &Corr.  &student \\
  &\omit &\omit  &\omit &\omit &\omit &\omit &coeff. &t$^6$ \\
\hline
\noalign{\smallskip}
$\DM$  &$\EM_\mathrm{p}$ &bis$^1$  &\multicolumn{2}{c}{$0.13^{+0.04}_{-0.03}$}
      &1.15&0.07   &$0.90\pm 0.08$   &11.8 \\
\noalign{\smallskip}
$\DM\cos|b|$  &$\EM_\mathrm{p}\cos|b|$   &"  &\multicolumn{2}{c}{$0.13^{+0.03}_{-0.02}$}
      &1.15&0.05   &$0.93\pm 0.06$   &14.7 \\
\noalign{\smallskip}
$\DM\sin|b|$  &$\EM_\mathrm{p}\sin|b|$   &"  &0.21&0.04  &0.98&0.09 &$0.82\pm 0.10$
      &\pheins 8.3 \\
\noalign{\smallskip}
$D$  &$\avnel$  &lin$^2$  &0.018&0.002   &\multicolumn{2}{c}{$-(2\pm6)10^{-4}$}
      &$0.52\pm 0.15$   &\pheins 3.5 \\
\noalign{\smallskip}
$|z|$   &$\avnel$  &exp$^3$  &0.019&0.002  &\multicolumn{2}{c}{$-3^{+1}_{-3}$}
     &$0.72\pm 0.12$    &\pheins 5.9 \\
\noalign{\smallskip}
$|z|$  &$\langle\nel^2\rangle$  &"  &0.0045&0.0007  &\multicolumn{2}{c}{$-1.4^{+0.3}_{-0.6}$}
     &$0.77\pm 0.11$   &\pheins 6.9 \\
\noalign{\smallskip}
$|z|$  &$\Nec$   &"   &\multicolumn{2}{c}{$0.24^{+0.04}_{-0.03}$}
     &\multicolumn{2}{c}{$-2.6^{+1.0}_{-4.1}$} &$0.76\pm 0.12$  &\pheins 6.6 \\
\noalign{\smallskip}
$|z|$  &$\Filfac$ &"  &\multicolumn{2}{c}{$0.077^{+0.020}_{-0.016}$}
     &\multicolumn{2}{c}{$[+23]^5$}   &$0.63\pm 0.14$  &\pheins 4.6 \\
\noalign{\smallskip}
$|z|$  &$L_\mathrm{e}\sin|b|$  &pow$^4$   &80&20  &1.06&0.15 &$0.83\pm 0.10$ &\pheins 8.5 \\
\noalign{\smallskip}
$\Nec$  &$\Filfac$  &"  &\multicolumn{2}{c}{$0.011^{+0.003}_{-0.002}$} &$-$1.20&0.13
       &$0.88\pm 0.09$   &10.2 \\
\noalign{\smallskip}
$\Nec$   &$L_\mathrm{e}$  &"   &\multicolumn{2}{c}{$24^{+20}_{-11}$} &$-$0.85&0.35
       &$0.75\pm 0.12$   &\pheins 6.5 \\
\noalign{\smallskip}
$\Nec$   &$L_\mathrm{e}\sin|b|$  &"  &2&1   &$-$1.6&0.3 &$0.82\pm 0.10$ &\pheins 8.0 \\
\hline
\noalign{\smallskip}
\multicolumn{9}{l}{Units: $\DM$ in $\cmcube\pc$, $\EM$ and $\EM_\mathrm{p}$ in $\cmsix\pc$,
  $\avnel$ and $\Nec$ in $\cmcube$, $\langle\nel^2\rangle$ in $\cmsix$,}\\
\multicolumn{9}{l}{ $|z|$ in kpc, $L_\mathrm{e}$ in pc }\\
\multicolumn{9}{l}{1) Bisector fit to powerlaw $Y=aX^\mathrm{b}$ in $\log X-\log Y$ plane.}\\
\multicolumn{9}{l}{2) Linear fit; regression line of $Y$ on $X$, where $Y=a+bX$.}\\
\multicolumn{9}{l}{3) Exponential fit; regression line of $\ln Y$ on $|z|$ with slope $1/b$,
   where $b$ is the scale}\\
\multicolumn{9}{l}{height $H$ in kpc.}\\
\multicolumn{9}{l}{4) Powerlaw fit; regression line of $\log Y$ on $\log X$.}\\
\multicolumn{9}{l}{5) Undetermined: errors $> 1/b$.}\\
\multicolumn{9}{l}{6) Student test: for a sample of 34 pulsars the correlation is
   significant at the $3\sigma$ level if $t > 3.3$.}\\
\multicolumn{9}{l}{All fits are ordinary least-squares fits with errors of one standard
    deviation.}\\
\end{tabular}
\end{table*}
%--------------------------------------------------------------------------------

\subsection{Dependence of $\avnel$, $\langle\nel^2\rangle$, $\Nec$ and
$\Filfac$ on height}

\begin{figure} %fig 4
\includegraphics[bb = 70 52 582 762,angle=270,width=8.8cm]{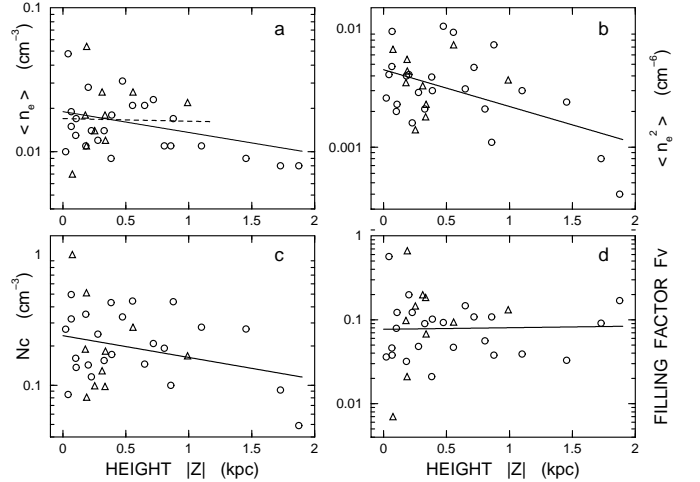}
\caption{Electron densities averaged along the line of sight and average
volume filling factor as function of height above the Galactic plane,
$|z|$.
{\it a)}\ Average density $\avnel = \DM \sin|b|/|z|$. The dashed line
is the exponential fit to the data at $|z| < 1.2\kpc$.
{\it b)}\  Average of the square of the density $\langle\nel^2\rangle =
\EM_\mathrm{p}\sin|b|/|z|$.
{\it c)}\  Mean density in clouds $\Nec = \EM_\mathrm{p}/\DM$.
{\it d)}\  Average volume filling factor $\Filfac =
\DM^2/(\EM_\mathrm{p} D)$.
Full lines indicate the exponential fits given in Table~\ref{tab:2}.
Circles: pulsars with $\Delta D/D < 0.2$; triangles: pulsars with $0.2
< \Delta D/D < 0.5$. }
\label{fig:4}
\end{figure}

Figure~\ref{fig:4} presents the various densities and $\Filfac$ as a
function of $|z|$ in the $\log Y - |z|$ plane. Although these variables
represent averages along the line of sight, we approximated their
$|z|$-distributions by exponentials with scale height $H$. The fits
are listed in Table~\ref{tab:2}.

The distribution of $\avnel(z)$ in Fig.~\ref{fig:4}a is effectively
constant up to $|z| \simeq 1\kpc$ (dashed line), a well known fact first
noted by Weisberg et al. (\cite{weisberg+80}). Beyond this height
$\avnel$ decreases. A fit to all points (full line) gives a mean value
at the midplane of $\avnel_0 = 0.019\pm 0.02\cmcube$, in fair
agreement with the expected value $\avnel_0 = \nel(0) = 0.023 \pm
0.004\cmcube$ derived in Sect.~3. The scale height is less well
determined, though. The spread in the data clearly decreases away from
the plane, as was also noted by BMM. This is not due to longer
pathlengths, because the spread in the distribution projected along the
plane remains constant (not shown). It just shows that the variety in
electron density is larger near the Galactic plane than away from the
plane.

The other panels in Fig.~\ref{fig:4} show the dependencies of
$\langle\nel^2\rangle$, $\Nec$ and $\Filfac$ on $|z|$. All quantities vary
considerably about the fitted lines. The mean density in clouds slowly
decreases from $0.24\cmcube$ at $|z| = 0\kpc$ to $0.16\cmcube$ at
$|z| = 1\kpc$, but at all heights values between $0.08\cmcube$ and
$0.5\cmcube$ occur. The midplane values of the exponential fits in
Table~\ref{tab:2} agree to within errors with those of $\nel^2(0)$,
$n_\mathrm{c0}$ and $f_0$ derived in Sect.~3 and the scale height $H$ of
$\langle\nel^2\rangle$ is indeed about half that of $\avnel$ and $\Nec$.
This agreement shows that the data are internally consistent.

The slight increase in $\Filfac(z)$ in Fig.~\ref{fig:4}d is
statisically insignificant, thus $\Filfac(z)$ is effectively a
constant of about $0.08 \pm 0.02$. This differs from the clear increase
in $\Filfac(z)$ found by BMM for a much larger sample of pulsars. The
spread in $\Filfac$ seems largest near the Galactic plane, but this
impression is mainly due to three extreme points of nearby pulsars at
low $|z|$: two near $\Filfac = 0.6$ and one near $\Filfac = 0.07$. At
$|z|> 0.3\kpc$ the spread stays within a factor of 4.

\begin{figure} %fig 5
\includegraphics[bb = 103 72 553 663,angle=270,width=8.8cm]{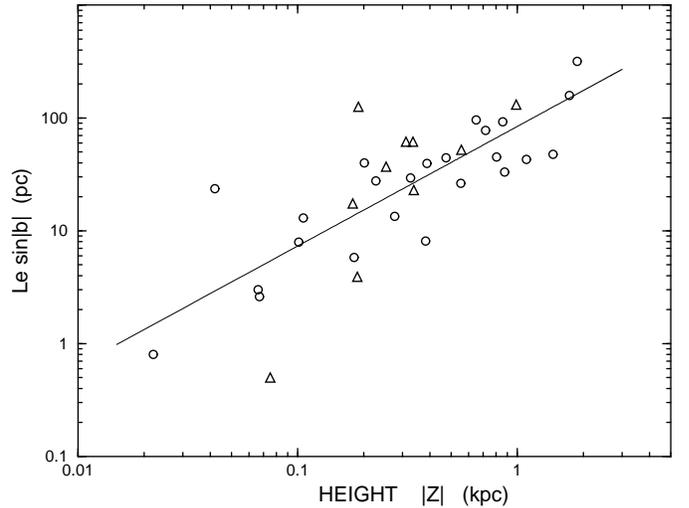}
\caption{Dependence of the total pathlength through ionized regions
perpendicular to the Galactic plane, $L_\mathrm{e} \sin|b|$, on height
above the plane. Circles: pulsars with $\Delta D/D < 0.2$;
triangles: pulsars with $0.2 < \Delta D/D < 0.5$. Full line: powerlaw fit
given in Table~\ref{tab:2}. }
\label{fig:5}
\end{figure}

If the mean filling fraction remains about constant along the
line of sight, the total pathlength through the ionized regions
will increase linearly with distance. $L_\mathrm{e} \sin|b|(z)$ indeed
increases nearly linearly from about $7\pc$ towards $|z| = 0.1\kpc$ to
about $80\pc$ towards $|z| = 1\kpc$ (see Fig.~\ref{fig:5} and
Table~\ref{tab:2}).

\subsection{Dependence of $\Filfac$ on $\Nec$}

One of the indications for turbulence in the ISM is an
inverse relationship between filling factor and mean cloud
density. This is not only expected on theoretical grounds (Fleck\
\cite{fleck96}; Elmegreen\ \cite{elmegreen99}), but has also been
obtained from simulations (Elmegreen\ \cite{elmegreen97}; Kowal \&
Lazarian\ \cite{kowal+07}). Observational evidence, however, is scarce
(see Sect.~1).

\begin{figure} %fig 6
\includegraphics[bb = 98 72 553 663,angle=270,width=8.8cm]{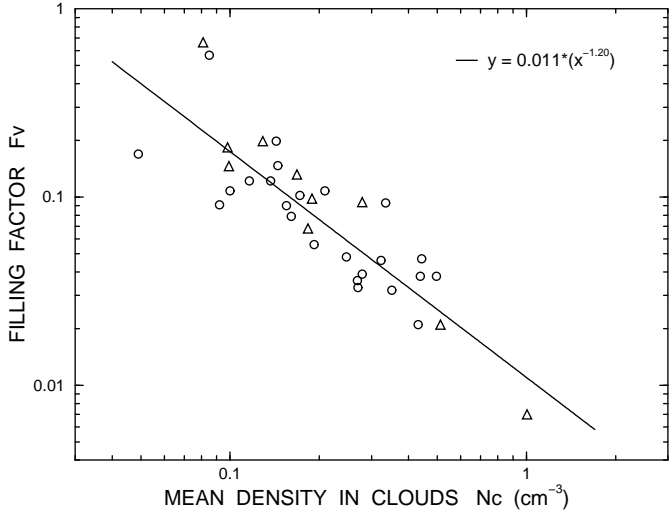}
\caption{Dependence of the average volume filling factor, $\Filfac$, on
the mean density in clouds, $\Nec$, for the sample of 34 pulsars in the
solar neighbourhood. Circles: pulsars with $\Delta D/D < 0.2$;
triangles: pulsars with $0.2 < \Delta D/D < 0.5$. The full line
shows the powerlaw fit given in Table~\ref{tab:2}. }
\label{fig:6}
\end{figure}

Since $\avnel$ is about constant (Fig.~\ref{fig:4}a) and $\avnel =
\Filfac \Nec$, we expect an inverse relationship between $\Filfac$ and
$\Nec$. We present this relationship for our sample
in Fig.~\ref{fig:6}. The correlation between $\Filfac$ and $\Nec$ is
very good and indeed nearly inverse: the power-law fit yields
$\Filfac (\Nec ) = (0.011\pm 0.003) \Nec^{-1.20\pm 0.13}$ with a
correlation coefficient of $0.88\pm 0.09$, and covers about 1.5 decade
in $\Nec\ (0.05-1 \cmcube)$ and $\Filfac\ (0.4-0.01)$. We compare this
result with earlier determinations and discuss its meaning for the
density structure in Sect.~5.3.

\begin{figure} %fig 7
\includegraphics[bb = 242 34 582 753,angle=270,width=8.8cm]{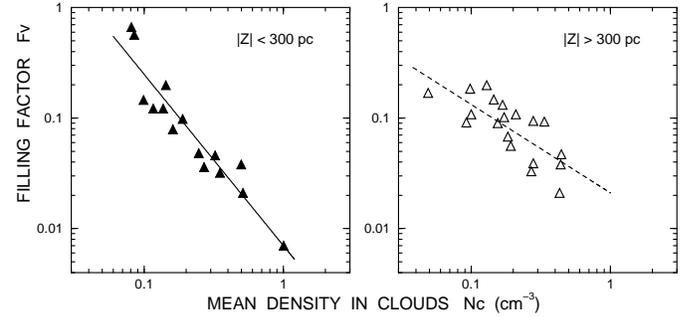}
\caption{Dependence of the relation $\Filfac (\Nec)$ on distance to the
Galactic plane $|z|$. Left: $|z| < 0.3\kpc$; right: $|z| > 0.3\kpc$.
The lines are powerlaw fits given in Table~\ref{tab:3}. }
\label{fig:7}
\end{figure}

As $\Filfac(\Nec)$ is well defined, we looked for variations with
longitude, distance along the plane and height above the
plane that may indicate variations in structure in the DIG.
We only found a dependence on height: at $|z| < 0.3\kpc$ the
relationship is considerably steeper and the filling factor
for $\Nec = 1\cmcube$ considerably smaller than at $|z| > 0.3\kpc$
(see Fig.~\ref{fig:7} and Table~\ref{tab:3}). The two lines cover nearly
the same range in $\Nec$ and cross near $\Nec = 0.15\cmcube$, but at
$|z| > 0.3\kpc$ the spread in the data is larger than at lower $|z|$.
On examination of the much larger sample of BMM, it appears to show the
same trend (see Table~\ref{tab:3}). The large number of pulsars even
permits a division of the lower $|z|$--interval yielding exponents of
$-1.6\pm 0.2$ for $|z| < 0.2\kpc$ and $-1.1\pm 0.1$ for $0.2 < |z| <
0.3\kpc$. It seems that in the thin Galactic disc $\Filfac$ depends
more strongly on $\Nec$ than above $|z| = 0.2\kpc$, with an exponent
significantly smaller than $-1$. This indicates that the structure of
the DIG in the thin disc differs from that away from the plane. This
may not be surprising. In the thin disc the activities of stellar winds,
classical \ion{H}{II} regions and supernova remnants largely determine
the structure of the DIG. They may change or destroy the structure of
the turbulence that is typical for more quiet regions away from the
Galactic plane. Interestingly, Cordes \& Lazio (\cite{cordes+lazio02})
derived a higher fluctuation factor for the thin disc than for the
thick disc in their model of the electron density distribution.
There is also evidence from measurements of interstellar
scintillation that different types of turbulent spectra exist in
different regions of the Galaxy (Shishov et al.\ \cite{shishov+03}).

\begin{table*}
\caption{$|z|$-dependence of the relation $\Filfac(\Nec) = a \Nec^b$}
\label{tab:3}
\centering
\begin{tabular}{l c r@{$\pm$}l r@{$\pm$}l c r@{$\pm$}l c}
\hline\hline
Sample  &$|z|$   &\multicolumn{2}{c}{$a$} &\multicolumn{2}{c}{$b$}
  & N  &\multicolumn{2}{c}{Corr.}  &student \\
  &$[\hbox{kpc}]$ &\omit&\omit &\omit&\omit &
  &\multicolumn{2}{c}{coeff.} &t$^1$ \\
\hline
This  &$0.0-0.3$  &0.0070&0.0018  &$-$1.55&0.15  &\pheins 15  &0.94&0.09 &10.2 \\
work  &$0.3-2.0$  &0.021\pheins &0.006 &$-$0.80&0.17   &\pheins 19  &0.80&0.14
      &\pheins 5.6 \\
\noalign{\smallskip}
BMM   &$0.0-0.2$  &0.0078&0.0020  &$-$1.62&0.16  &\pheins 11  &0.97&0.08  &12.7 \\
      &$0.2-0.3$  &0.0183&0.0037  &$-$1.11&0.12  &\pheins 24  &0.91&0.09  &10.1 \\
\noalign{\smallskip}
      &$0.0-0.3$  &0.0137&0.0023  &$-$1.29&0.10  &\pheins 35  &0.92&0.07  &13.9 \\
      &$0.3-2.0$  &0.0187&0.0014  &$-$1.04&0.03  &122         &0.95&0.03  &31.9 \\
\hline
\noalign{\smallskip}
\multicolumn{10}{l}{1) student test: for a sample of $N \ge 11$ pulsars the
correlation is significant at the} \\
\multicolumn{10}{l}{$3\sigma$ level if $t > 4.0$} \\
\end{tabular}
\end{table*}
%--------------------------------------------------------------------------------

\section{Discussion}

\subsection{Comparison with BMM 2006}

It is interesting to compare our results with those of BMM who
took the distances to the pulsars from the NE2001 model of Cordes
\& Lazio (\cite{cordes+lazio02}) and used an average correction for the
absorption of the H$\alpha$ emission that only depends on latitude.
Applying their method to our sample, we recalculated the main
relationships using a scale height of $0.47\kpc$ to correct for
H$\alpha$ emission from behind the pulsar. Table~\ref{tab:4} shows that
the resulting relations are nearly identical to those derived in
Sect.~4, but with larger errors. This means that the less accurate
distances and absorption corrections used by BMM have \emph{not}
influenced their statistical results. Therefore, we may compare results
from our small sample with those from the much larger sample analyzed
by BMM with confidence.

\begin{table*}
\caption{Comparison with BMM for our sample of 34 pulsars}
\label{tab:4}
\centering
\begin{tabular}{l@{ $=$ }l r@{$\pm$}l r@{$\pm$}l r@{$\pm$}l r@{$\pm$}l}
\hline\hline
\multicolumn{2}{c}{Relation}  &\multicolumn{4}{c}{Our data} &\multicolumn{4}{c}{BMM} \\
 \omit&\omit &\multicolumn{2}{c}{$a$}  &\multicolumn{2}{c}{$b$}
 &\multicolumn{2}{c}{$a$} &\multicolumn{2}{c}{$b$} \\
\hline
\noalign{\smallskip}
$\langle\nel^2\rangle$ &$a\, e^{-|z|/b}$  &0.0045&0.0007  &\multicolumn{2}{c}{$1.4^{+0.6}_{-0.3}$}
   &0.0053&0.0011 &\multicolumn{2}{c}{$1.4^{+1.5}_{-0.5}$} \\
\noalign{\smallskip}
$\Nec (z)$ &$a\, e^{-|z|/b}$  &0.24&0.04  &\multicolumn{2}{c}{$2.6^{+4.1}_{1.0}$} &0.26&0.06
   &\multicolumn{2}{c}{$2^{+10}_{-1}$} \\
\noalign{\smallskip}
$\Filfac (z)$ &$a\, e^{|z|/b}$  &0.077&0.018  &\multicolumn{2}{c}{$[23]^1$}  &0.077&0.023
    &\multicolumn{2}{c}{$[5]^1$} \\
\noalign{\smallskip}
$\Filfac (\Nec)$   &$a\, \Nec^b$  &0.011&0.003 &$-$1.20&0.13  &0.014&0.003 &$-$1.18&0.11 \\
\hline
\noalign{\smallskip}
\multicolumn{10}{l}{1) Undetermined: $1/b$ smaller than errors} \\
\end{tabular}
\end{table*}
%--------------------------------------------------------------------------------

A striking difference between the two samples is the behaviour
of $\Filfac(z)$. Where BMM found a significant increase of the
filling factor to about 0.2 towards $|z| = 1\kpc$, $\Filfac(z)$
derived from our sample remains constant within the errors at a value
of about 0.08. This difference must be due to the different locations
in the Galaxy of the pulsars in the two samples. For example, 17 out
of the 34 pulsars (50\%) in our sample are within $1\kpc$ from
the Sun, whereas only 44 out of 157 pulsars (28\%) of the BMM
sample are within this distance. Since the ISM is a highly
variable medium, differences between samples are to be expected.
Hydromagnetic simulations of the evolution of the ISM by de Avillez
\& Breitschwerdt (\cite{avillez+breit05}) may illustrate this point.
Their Fig.~2a shows the density variation in a cut of $1\kpc$ along and
$10\kpc$ perpendicular to the Galactic plane. Although the filling
fraction of the DIG generally increases with height, there are also
regions where it remains constant or even decreases away from the plane.

\subsection{Extent of ionized regions}

The total pathlength through ionized regions perpendicular to the
Galactic plane increases nearly linearly from about $7\pc$
towards $|z| = 100\pc$ to about $40\pc$ towards $|z| = 500\pc$ and
about $80\pc$ towards $|z|= 1\kpc$ (Fig.~\ref{fig:5}).

\begin{figure} %fig 8
\includegraphics[bb = 103 72 559 663,angle=270,width=8.8cm]{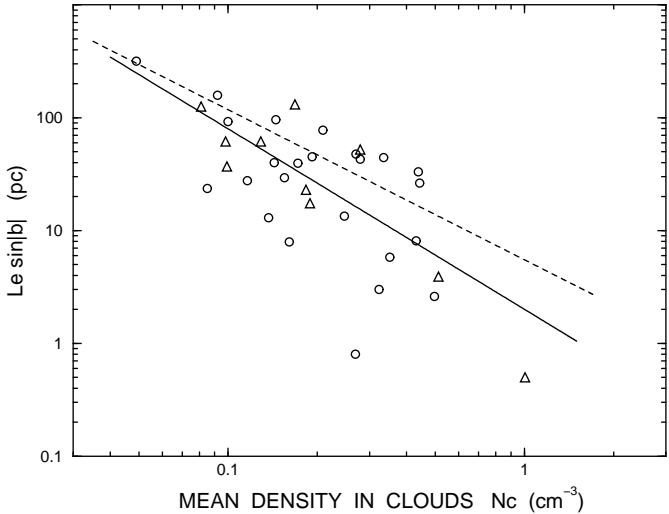}
\caption{Dependence of the total pathlength through ionized regions
perpendicular to the Galactic plane, $L_\mathrm{e} \sin|b|$, on the mean
density in clouds, $\Nec$. Circles: pulsars with $\Delta D/D < 0.2$;
triangles: pulsars with $0.2 < \Delta D/D < 0.5$. Full line:
powerlaw fit given in Table~\ref{tab:2}; dashed line: powerlaw fit to
the sample of 157 pulsars of BMM, $L_\mathrm{e} \sin|b|(\Nec) =
(5.5\pm 0.8) \Nec^{-1.32\pm 0.07}$. }
\label{fig:8}
\end{figure}

In the ISM dense regions are usually smaller than less dense
regions, which also applies to the DIG (see Fig.~\ref{fig:8}). Although
the distribution of $L_\mathrm{e} \sin|b|$ with $\Nec$ shows
considerable spread, it is well fitted by a power law (see
Table~\ref{tab:2}). Regions of density $\Nec = 1\cmcube$  together
occupy about $2\pc$ in the $|z|$-direction and those of $\Nec =
0.05\cmcube$ about $300\pc$.

We calculated the same relationship for the BMM sample which gives a
somewhat larger extent for the higher densities (see Fig.~\ref{fig:8}).
This power law is close to the one indicated by the distribution of 192
pulsars presented by Hill et al. (\cite{hill+07}). However, these
authors did not apply any correction to the observed emission measures,
which leads to a flattening of the distribution and a shift to higher
values of $\Nec$ for the same pathlength (see Fig.~5 in BMM).

\subsection{The $\Filfac$--$\Nec$ relation and density structure}

In spite of our small sample of pulsars, the nearly inverse
correlation between $\Filfac$ and $\Nec$ shown in Fig.~\ref{fig:6} is
very tight, reaching a correlation coefficient of $0.88\pm 0.09$ (see
Table~\ref{tab:2}). It does not change with distance along the plane
and, as BMM explained, is insensitive to errors in emission measure.
The near constancy of $\avnel = \Filfac \Nec$ along the line of sight
(see $D- \avnel$ fit in Table~\ref{tab:2}), first noted by Weisberg et
al. (\cite{weisberg+80}), was an early indication of an inverse
relationship between $\Filfac$ and $\Nec$.

\begin{figure} %fig 9
\includegraphics[bb = 98 56 554 669,angle=270,width=8.8cm]{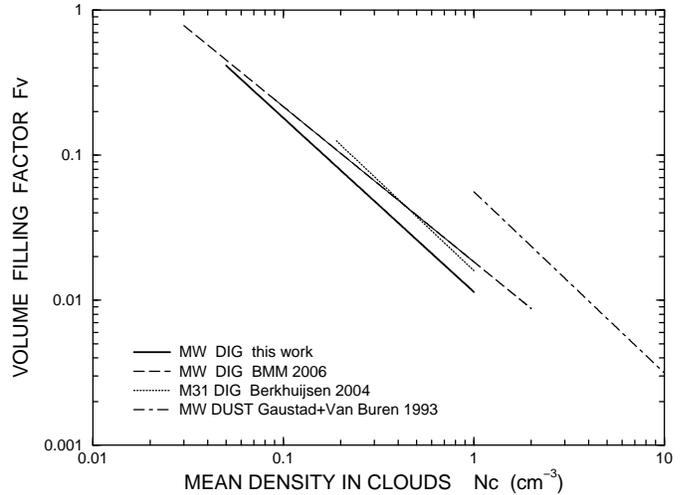}
\caption{Dependence of the volume filling factor, $\Filfac$, on mean
density in clouds, $\Nec$, as derived by different authors.
Full line: this work; dashed line: dependence for the DIG within about
$3\kpc$ from the Sun (BMM); dotted line: dependence for the DIG in the
bright emission ring in M31 (Berkhuijsen\ \cite{elly04});
dash-dot line: dependence for clouds of diffuse dust within $400\pc$
from the Sun (Gaustad \& Van Buren\ \cite{gaustad+93}). The parameters
of the lines are given in Table~\ref{tab:5}. }
\label{fig:9}
\end{figure}

We first compare our $\Filfac$--$\Nec$ relation with earlier
determinations available in the literature, and then discuss what it
tells us about the density structure in the DIG.

\subsubsection{Comparison with other data}

Figure~\ref{fig:9} and Table~\ref{tab:5} show our $\Filfac$--$\Nec$
relation together with earlier determinations. For the same mean
density in clouds, we find somewhat smaller filling factors than BMM
which agree with the lower points in their distribution.
The smaller errors in the BMM relation are not only due to the larger
sample, but also to the smooth distribution of $\avnel = \DM /D$ of the
electron density model of Cordes \& Lazio (\cite{cordes+lazio02}) used
by BMM. Berkhuijsen (\cite{elly04}) obtained the $\Filfac$--$\Nec$
relation for the DIG in the galaxy M\,31 from a comparison of rotation
measures of the polarized continuum emission (giving $\avnel$ with
magnetic field strengths of Fletcher et al.\ \cite{fletcher+04})
and thermal radio emission (giving $\langle\nel^2\rangle$). These data
refer to the DIG in the bright emission ring in M\,31. The good
agreement between the M\,31 and the MW results suggests that the inverse
correlation between $\Filfac$ and $\Nec$ is generally valid in the DIG
in galaxies.

\begin{table*}
\caption{Comparison of $\Filfac -\Nec$ relations $\Filfac (\Nec) = a\, \Nec^b$}
\label{tab:5}
\centering
\begin{tabular}{l r@{$\pm$}l r@{$\pm$}l c c c}
\hline\hline
Sample  &\multicolumn{2}{c}{$a$} &\multicolumn{2}{c}{$b$} &Range $\Nec$  &Range $\Filfac$ &Ref. \\
  &\omit&\omit &\omit&\omit &$[\cmcube]$ \\
\hline
DIG \\
MW\quad N= 34 &0.0114&0.0025  &$-$1.20&0.13  &$0.05-1$  &$0.4\pheins -0.011$ &1 \\
MW\quad N=157 &0.0184&0.0011  &$-$1.07&0.03  &$0.03-2$  &$0.8\pheins -0.009$ &2 \\
M31           &0.016 &0.004   &$-$1.24&0.30  &$0.19-1$  &$0.13-0.016$        &3 \\
\noalign{\smallskip}
DIFFUSE DUST \\
$D < 400\pc$  &0.056 &0.020   &$-$1.25&0.40  &$1-10$    &$0.06-0.003$        &4 \\
\hline
\noalign{\smallskip}
\multicolumn{8}{l}{References. (1) This work ; (2) Berkhuijsen, Mitra, \& M\"uller\
\cite{elly+06};} \\
\multicolumn{8}{l}{(3) Berkhuijsen\ \cite{elly04}; (4) Gaustad \& Van Buren\
\cite{gaustad+93}} \\
\end{tabular}
\end{table*}
%--------------------------------------------------------------------------------

Roshi \& Anantharamaiah (\cite{roshi+01}) observed radio recombination
lines from the inner Galaxy at low latitudes. They derived filling
factors of $\le 0.01$ for extended regions of diffuse ionized gas with
densities of $1-10\cmcube$, which agree well with the extension of
the curves for the DIG in Fig.~\ref{fig:9}.

Thus the inverse correlation between $\Filfac$ and $\Nec$ at least holds for
the density range $0.03-10\cmcube$. Cordes et al. (\cite{cordes+85})
estimated a filling factor of $10^{-4.0\pm0.3}$ for clumps of about $1\pc$
size causing scattering of pulsar signals. Our data predict this filling
factor for clumps of density $\Nec = 50\cmcube$ and size $L_\mathrm{e}/m
= 0.9\pc$, where $m$ is the number of clumps on the line of sight. The
relation of BMM gives clumps of $\Nec = 100\cmcube$ for the same
filling factor and sizes. So these clump properties also seem to follow
the $\Filfac$--$\Nec$ relations derived for the DIG.

Furthermore, Gaustad \& Van Buren (\cite{gaustad+93}) obtained a
similar relationship for clouds of diffuse dust within $400\pc$ from the
Sun, with mean densities of $1-10\cmcube$. Figure~\ref{fig:9} shows that
these clouds have about 3 times higher filling factors than ionized
clouds of the same mean density. Interestingly, their data agree very
well with those of Pynzar (\cite{pynzar93}, also shown in Berkhuijsen\
\cite{elly98}) derived from recombination lines.

\subsubsection{Density structure of the DIG }
The general validity of the inverse $\Filfac$--$\Nec$ relation
indicates that it describes a basic property of the DIG, possibly even
of the entire ISM (Berkhuijsen\ \cite{elly99}).
At least two mechanisms could be responsible: thermal pressure
equilibrium and turbulence causing a fractal density structure.

Thermal pressure equilibrium leads to an inverse $\Filfac$--$\Nec$
relation if it were widespread, but it seems only locally valid.
In the MW large fractions of gas are observed in unstable regimes
and have also been found in simulations of a turbulent ISM.
Thermal pressure equilibrium of clouds appears to be of minor
importance in the presence of turbulence (see Elmegreen \& Scalo\
\cite{elmegreen+04}, and references therein).

Elmegreen (\cite{elmegreen98}, \cite{elmegreen99}) discussed the
properties of diffuse ionized gas in a pervasive fractal ISM. One of
these properties is an inverse correlation between volume filling
factor and gas density, which is expected on theoretical grounds (Fleck\
\cite{fleck96}) as well as from simulations (Elmegreen\ \cite{elmegreen97};
Kowal \& Lazarian\ \cite{kowal+07}). As this leads to a constant
average density along the line of sight, the observed near constancy of
$\avnel$ with $\Nec$ and $\Filfac$ inversely varying over 1.5 decade
may indicate that the density distribution of the DIG is fractal.
Please note that the near constancy of $\avnel$ alone  should only be
regarded as a first indication, because the fits in Table~\ref{tab:2}
show that the statistical significance of the near constancy of
$\avnel$ is much lower than that of the nearly inverse
$\Filfac$--$\Nec$ relation.

The fractal medium is characterized by filamentary, clumpy structures
that tend to cluster together, with holes inside and large voids
between them. The filling factor along the line of sight is determined
by the outer regions of the filaments, i.e. by the largest scales,
because the small, dense clumps in the inner parts hardly contribute.
The voids occupy more space than the filaments, thus the filling factor
of the filaments is small. The values of $\Filfac \simeq 0.1$
(Sect.~4.1) and $\Filfac \simeq 0.2$ (BMM) obtained when looking
perpendicular to the Galactic plane towards $|z|=1\kpc$ are consistent
with this picture.

The tendency of filaments to cluster into larger complexes reduces
the number of fractal clouds, $N_\mathrm{f}$, along the line of sight.
Elmegreen (\cite{elmegreen98}) estimated $N_\mathrm{f}=3/{\rm kpc}$
locally. For $\Filfac=0.1$ the total pathlength through the ionized
regions is 100~pc/kpc, giving a line of sight through one fractal
complex of about $30\pc$ and a mean density $\Nec$ of $0.15\cmcube$
(from Fig.~\ref{fig:6}). Using the results of BMM gives $\Filfac=0.2$, a
fractal complex of about $70\pc$ along the line of sight and $\Nec =
0.10\cmcube$. Interestingly, these pathlengths are in the range of
10--100~pc that Ohno \& Shibata (\cite{ohno+93}) estimated for
the cells of ionized gas causing $\DM$ and rotation measure $\RM$
towards pairs of pulsars close on the sky.

According to Elmegreen (\cite{elmegreen98}), pervasive fractal
structure only develops in regions outside the influence of SN shells,
sites of star formation, chimneys etc. that are concentrated near the
Galactic plane. This may explain our result of Sect.~4.2 that the
exponent of the $\Filfac$--$\Nec$ relation is significantly smaller in
the thin disc ($|z|<300\pc$) than further from the plane where it is
close to $-1$ (see Table~\ref{tab:3} and Fig.~\ref{fig:7}).

We conclude that our results on the mean electron densities and
volume filling factors of the DIG outside the thin disc are
consistent with a fractal ionized medium caused by turbulence,
while the structure in the thin disc is dominated by the effects
of SN shocks, stellar winds, star formation and other forces. The
difference in density structure between these two regimes could be
further analyzed when more pulsars with measured distance become
available.

Recently, further observational evidence for turbulent structure of the
DIG and of the diffuse atomic gas has been presented. Hill et al.
(\cite{hill+08}) found that the probability distribution function (PDF)
of the emission measures observed perpendicular to the Galactic plane at
$|b| > 10\degr$ is lognormal as is expected for a turbulent medium from
MHD simulations of the ISM (Elmegreen \& Scalo\ \cite{elmegreen+04}).
Furthermore, Berkhuijsen \& Fletcher (\cite{elly+08}) showed that the
PDFs of the quantities $\avnel$, $\langle\nel^2 \rangle$, $\Nec$ and
$\Filfac$ derived above as well as that of the mean \ion{H}{I} density
along the line of sight towards stars, $\langle n_\mathrm{HI}\rangle$,
are lognormal, consistent with a turbulent origin of density structure
in the diffuse gas.

\section{Summary}

We have used pulsar dispersion measures (Manchester et al.\
\cite{manchester+05}) and extinction-corrected emission measures
(Finkbeiner\ \cite{finkbeiner03}; Dickinson et al.\ \cite{dickinson+03})
towards 38 pulsars with known distances for a statistical study of
several parameters of the diffuse ionized gas (DIG) in the solar
neighbourhood. The emission measures were also corrected for
contributions from beyond the pulsar distance. To avoid regions with
strong absorption ($> 1$~magnitude) and contributions from \ion{H}{II}
regions most pulsars in the sample are at Galactic latitudes $|b| >
5\degr$. The statistical results are collected in Tables~\ref{tab:2} and
\ref{tab:3}.

\medskip

Our main conclusions are:
\begin{enumerate}
\item From the scaling of dispersion measures with distance perpendicular
   to the Galactic plane, we find a scale height of the ionized layer
   of $0.93\pm 0.13\kpc$ and an electron density at the midplane of
   $0.023\pm 0.004\cmcube$, in good agreement with earlier
   determinations.

\item Dispersion measure $\DM$ and corrected emission measure
   $\EM_\mathrm{p}$ are well correlated, indicating that they probe the
   same ionized regions. We may then combine them to derive the average
   densities along the line of sight, $\avnel$ and
   $\langle\nel^2\rangle$, the mean electron density in clouds
   in the line of sight, $\Nec$, the volume filling factor of these
   clouds, $\Filfac$, and the total pathlength through the ionized
   regions, $L_\mathrm{e}$.

\item The total extent of the ionized regions perpendicular to the
   Galactic plane increases linearly reaching about $80\pc$ towards
   $|z| = 1\kpc$.

\item The filling factor $\Filfac$ remains essentially constant with
   $|z|$ at a mean value of $0.08\pm 0.02$. Whether $\Filfac(z)$ is
   constant or systematically increases with $|z|$ depends on the
   regions in the Galaxy probed by the pulsar sample.

\item The average electron density $\avnel$ is about constant with a
   spread of a factor $\la 2$ about the mean value of $0.018 \pm
   0.002\cmcube$. Since $\avnel = \Filfac \Nec$, an inverse
   relationship between $\Filfac$ and $\Nec$ is expected.

\item We derived the relation $\Filfac (\Nec) = (0.011\pm 0.003)
   \Nec^{-1.20\pm 0.13}$, which covers about 1.5 decade in $\Nec$
   $(0.05-1\cmcube)$ and $\Filfac$ $(0.4-0.01)$ (see Fig.~\ref{fig:6}).
   This is in good agreement with earlier determinations of the
   $\Filfac$--$\Nec$ relations for the DIG (see Fig.~\ref{fig:9}).

\item Near the Galactic plane the dependence of $\Filfac$ on $\Nec$
   is significantly stronger than away from the plane. This indicates that
   the thin disc has a different turbulent structure than regions away
   from the plane.

\item The inverse relationship between $\Filfac$ and $\Nec$, and hence
   also the near constancy of $\avnel$, are consistent with  a
   fractal density distribution in the DIG caused by turbulence which
   dominates the structure outside the thin disc.

\end{enumerate}

\begin{acknowledgements}
We thank Dr.~Rainer~Beck for careful reading of the manuscript and
useful suggestions, and the referee for helpful comments leading to
improvements in the manuscript.
\end{acknowledgements}

\end{document}